\pdfoutput=1

\documentclass[aps,twocolumn,amsmath,amssymb,nofootinbib,superscriptaddress,showkeys,prd,floatfix]{revtex4-1} 

\usepackage{epsfig}
\usepackage{slashed}
\usepackage{hyperref}

\newcommand{\vect}[1]{\boldsymbol{#1}}

\begin{document} 

\title{Double parton distribution of valence quarks in the pion in chiral quark models}

\author{Wojciech Broniowski}
\email{Wojciech.Broniowski@ifj.edu.pl}
\affiliation{Institute of Physics, Jan Kochanowski University, 25-406 Kielce, Poland}
\affiliation{H. Niewodnicza\'nski Institute of Nuclear Physics PAN, 31-342 Cracow, Poland}

\author{Enrique Ruiz Arriola}
\email{earriola@ugr.es}
\affiliation{Departamento de F\'isica At\'omica, Molecular y Nuclear and Instituto Carlos I de F\'{\i}sica Te\'orica y Computacional \\ 
Universidad de Granada, E-18071 Granada, Spain}

\date{31 December 2019}

\begin{abstract}
The valence double parton distribution of the pion is analyzed in the
framework of chiral quark models, where in the chiral limit
factorization between the longitudinal and transverse degrees of
freedom occurs. This feature leads, at the quark-model scale, to a
particularly simple distribution of the form $D(x_1,x_2, \vect{q})
= \delta(1-x_1-x_2) F(\vect{q})$, where $x_{1,2}$ are the longitudinal
momentum fractions carried the valence quark and antiquark and
$\vect{q}$ is their relative transverse momentum.  For
$\vect{q}=\vect{0}$ this result complies immediately to the 
Gaunt-Sterling sum rules.  The DGLAP evolution to higher scales is
carried out in terms of the Mellin moments.  We then explore its role
on the longitudinal correlation quantified with the ratio
of the double distribution to the product of single distributions,
$D(x_1,x_2, \vect{q}=\vect{0})/D(x_1)D(x_2)$. We point out that the ratios of moments 
$\langle x_1^n x_2^m \rangle / \langle x_1^n \rangle \langle
x_2^m \rangle $ are independent of the evolution, providing
particularly suitable measures to be tested in the upcoming lattice
simulations.  The transverse form factor $F(\vect{q})$ and its Fourier conjugate in 
the relative transverse coordinate $\vect{b}$ are obtained in
variants of the Nambu--Jona-Lasinio model with the spectral and Pauli-Villars regularizations.
The results are valid in the soft-momentum domain.
Interestingly, with the spectral regularization of the model, the effective cross section for the
double parton scattering of pions is exactly equal to the geometric cross
section, $\sigma_{\rm eff}=\pi \langle \vect{b}^2 \rangle$ and yields
about 20~mb. 
\end{abstract}


\maketitle

\section{Introduction \label{sec:intro}}

The pion, which according to the constituent quark model is composed
of a quark $q$ and antiquark $\bar q$ pair, enjoys being the would-be
Goldstone boson of the spontaneously broken chiral symmetry -- a
feature which explains its particularly low mass as compared to other
hadrons. There exist many observables which probe {\it indirectly} the
consequences of the pion being {\it both} a pseudo-Goldstone boson
{\it and} a composite $q \bar q$ state, indicating its extended
non-elementary structure at intermediate energies. These include the
electromagnetic and gravitational form factors (FF), transition form
factors, electromagnetic polarizabilities, as well as features of
their the mutual interactions, such as the Gasser-Leutwyler
coefficients. Likewise, for high-energy processes where hard-soft
factorization holds, the soft matrix elements within the pion state
yield the single parton distribution functions (sPDF), the
corresponding generalized parton distributions (GPD), transverse
momentum distributions (TMD), parton distribution amplitudes (PDA),
etc.

A strong motivation to investigate the double parton distribution
functions (dPDFs) in the pion comes from the more and more promising
prospects of the lattice QCD
studies~\cite{Bali:2018nde,Zimmermann:2017ctb}.  In this paper we
address the valence dPDF of the pion in the simplest covariant field-theoretic
approach to its structure, namely the Nambu--Jona-Lasinio (NJL)
model~\cite{Nambu:1961tp,Nambu:1961fr}, where the pion arises as a
Goldstone boson of the spontaneously broken chiral symmetry and as a
$q\bar q$ relativistic bound state of the Bethe-Salpeter
equation. This model is not renormalizable and needs a suitable
regularization. Here we use the Pauli-Villars (PV)
regularization which preserves chiral symmetry and gauge
invariance~\cite{Schuren:1991sc} (for a review see,
e.g.,~\cite{RuizArriola:2002wr}). We also explore the Spectral Quark
Model (SQM)~\cite{RuizArriola:2001rr,RuizArriola:2003bs}, where particularly simple
analytic formulas can obtained.

The story of dPDFs is rather old~\cite{Kuti:1971ph}, with first experimental traces of
the possible double parton scattering (DPS) in multi-jet events in
$pp$ collisions reported by the Axial Field Spectrometer Collaboration
at the CERN ISR~\cite{Akesson:1986iv} and in $p\bar{p}$ collisions by
the CDF Collaboration at Fermilab~\cite{Abe:1993rv,Abe:1997xk}.
With a renewed interest in the LHC era, several enlightening theoretical
works (see~\cite{Bartalini:2011jp,Snigirev:2011zz,Luszczak:2011zp,Manohar:2012jr,Manohar:2012pe}
and references therein)
preceded the measurement of $W$ production in association with two
jets by the ATLAS Collaboration~\cite{Aad:2013bjm}, which
provided a strong case for the need of DPS. Thereby dPDFs have become
more directly accessible to experimental scrutiny and quantitative
analysis (for reviews see, e.g.,~\cite{d'Enterria:2012qx} for
implications in production processes). For a recent state of affairs see~\cite{Bartalini:2017jkk}.

As any partonic quantity, dPDFs are scale dependent and their
evolution is dictated by a generalized form of the DGLAP
equations~\cite{Kirschner:1979im,Shelest:1982dg}, where gluon
radiation is explicitly implemented. In the case of the pion, the
correlation structure is particularly relevant, as it indicates to
what extent the quark and anti-quark are entangled in the
presence of additional gluons. While it will be hard to measure dPDF
for the pion experimentally, some of its features will become soon
available from lattice QCD~\cite{Bali:2018nde,Zimmermann:2017ctb},
thus offering a unique chance of testing the internal structure of the
pion and, more specifically, the correlations between its $q \bar q$
constituents. However, the quest for dPDFs is non-perturbative,
whereas the evolution only relates these quantities perturbatively at
different scales, saying nothing about their absolute determination at
a given scale. Hence the importance of non-perturbative modeling. 

Similarly to the previous studies of sPDFs, we {\it assume} that there
exists a reference scale $\mu_0$ where the pion is just a $q\bar q$
state. This is suggested by the observation that the momentum fraction
carried by the quarks at the scale $\mu= 4~{\rm GeV}$ is about
$42-44\%$~\cite{Sutton:1991ay,Gluck:1999xe} and that it increases according to LO and NLO DGLAP evolution with decreasing $\mu$. 
Then at about $\mu_0 \sim 320~{\rm MeV}$ the valence quark and antiquark carry all the 
momentum. While this is admittedly a low
scale, it has been checked that switching from LO to NLO evolution in
model calculations does not change more than $10\%$ the pion
sPDFs~\cite{Davidson:2001cc}, justifying the approach.

Chiral quark models offer a view of the pion as a relativistic $q\bar
q$ Goldstone boson, departing from the naive quark model. One of the
important features of these models is that they are constructed to
satisfy the gauge invariance and the chiral Ward-Takahashi
identities. This is important to recover the current algebra results
at low energies and to guarantee proper normalization of form factors
and PDFs. Moreover, the imposition of relativity via the
Bethe-Salpeter equation implements proper support for partonic
quantities.  Despite these strong constraints, there is still freedom
in enforcing proper regularization schemes. In this paper we focus for
definiteness on the NJL model with the PV
regularization~\cite{Schuren:1991sc,RuizArriola:2002wr} for which
sPDFs~\cite{Davidson:1994uv}, TMDs~\cite{Weigel:1999pc},
PDAs~\cite{RuizArriola:2002bp}, GPDs~\cite{Broniowski:2007si} have
been computed. We also explore the SQM, which is particularly simple
in the resulting formulas and offers a qualitatively good description
both at low and at higher
energies~\cite{RuizArriola:2003bs,RuizArriola:2003wi,Megias:2004uj}. These
variants of the NJL model have been used in the past to describe the
FF, PDFs, PDAs, GPDs, TMDs, comparing favorably both to experimental
data as well as available lattice results.

After some of the results presented here were
advertised~\cite{BW-ERA-LC2019}, a relevant preprint by Courtoy, Noguera, and
Scoppeta has been released~\cite{Courtoy:2019cxq}, addressing in a
comprehensive way the dPDF calculation in the NJL model.  Our results
confirm closely the basic findings~\cite{Courtoy:2019cxq}, despite different
regularization schemes.\footnote{Ref.~\cite{Courtoy:2019cxq} also uses a PV regulator (but at
difference from ours, see below) as well as a light-cone cut-off.}  Here we
also fully present the results of the QCD evolution for dPDF and
discuss in detail the issue of the longitudinal partonic correlation.

It is worth noting some other non-perturbative studies of dPDFs in quark
models for the nucleon, such as in the MIT bag~\cite{Chang:2012nw} or
in the constituent quark model model~\cite{Rinaldi:2013vpa}, also with
the light-front wave functions~\cite{Rinaldi:2014ddl,Rinaldi:2018zng}.
We stress that the longitudinal correlation features in these models
are qualitatively understood in the framework of the {\em valon}
model~\cite{Hwa:1980mv,Hwa:2002mv}, as shown in~\cite{Broniowski:2013xba,Broniowski:2016trx}.\footnote{In the
sPDF case, the longitudinal results for the pion are also consistent
with the simple valon model
picture~\cite{RuizArriola:1999hk,RuizArriola:2001rr}.}

The paper is organized as follows. After introducing the definitions
and our notation in Sec.~\ref{sec:defs}, we present the NJL model and
the calculation of the valence dPDF of the pion in
Sec.~\ref{sec:model}.  A recurrent issue in this kind of calculations
regards the interplay between regularization and positivity, a topic
which we address in Section~\ref{sec:reg-vs-pos}. The QCD DGLAP
evolution and the corresponding matching condition for the model are
discussed in Sec.~\ref{sec:evol}, with numerical results presented in
Sec.~\ref{sec:num}. In Sec.~\ref{sec:ratios} we present ratios of
moments of dPDF to product of moments of sPDFs, which are scale
independent for the valence distributions.  These ratios could most
directly be tested in lattice calculations. The transverse form factor
depends somewhat on the variant of the model, as shown in
Sec.~\ref{sec:trans}, where also a few related observables of interest
are evaluated and discussed. Finally, in Sec.~\ref{sec:concl} we come
to our summary.

\section{Basic definitions and properties \label{sec:defs}}

In this section we list for completeness (and to establish our
notation) the basic definitions.  The {\em spin-averaged} sPDF and
dPDF~(see \cite{Diehl:2010dr} and references therein) of a hadron with momentum $p$
involve forward matrix elements of parton bilinear operators, namely
\begin{eqnarray}
  && \hspace{-2mm} D_{j_1}(x_1) =   \int \frac{d z^-}{2\pi}\,
    e^{i x_1^{} z^- p_{}^+} \langle p |\,
   { \mathcal{O}_{j_1}(0,z)} \,| p \rangle
    \bigl|_{z^+ = 0\,, \vect{z}^{} = \vect{0}},  \nonumber \\
&&  \hspace{-2mm} D_{j_{1} j_{2}}(x_1,x_2,\vect{b})
 =  2 p^+ \int d y^-\,
        \frac{d z^-_1}{2\pi}\, \frac{d z^-_2}{2\pi}\;
          e^{i ( x_1^{} z_1^- + x_2^{} z_2^-) p_{}^+} \nonumber \\
 && \times
    \langle p |\, {\mathcal{O}_{j_1}(y,z_1)\, \mathcal{O}_{j_2}(0,z_2)}
    \,| p \rangle
    \bigl|_{z_1^+ = z_2^+ = y_{\phantom{1}}^+ = 0\,,
    \vect{z}_1^{} = \vect{z}_2^{} = \vect{0}},  \nonumber \\ \label{eq:defsd}
\end{eqnarray}
with $j_{1,2}$ indicating the parton species, the bold face denoting
the transverse vectors, and {$\vect{b}$ playing the role of the
transverse distance between the two partons, with $y=(y^+,y^-,\vect{b})$.  The light-cone
coordinates are defined as $a^\pm = (a^0 \pm a^3) /\sqrt{2}$.  The
longitudinal (i.e., ${}^+$) momentum fractions carried by the partons
are denoted customarily as $x_1$ and $x_2$.

For the quarks and antiquarks, considered in this paper, the bilocal color-singlet operators are
\begin{eqnarray}
&& \hspace{-2mm} \mathcal{O}_{q}(y, z) = \tfrac{1}{2}\, \bar{q} ( y
- \tfrac{z}{2}) \gamma^+ W( y - \tfrac{z}{2}, y + \tfrac{z}{2}) q ( y
+ \tfrac{z}{2} ), \nonumber \\
&& \hspace{-2mm} \mathcal{O}_{\bar{q}}(y, z) =
-\tfrac{1}{2}\, \bar{q} ( y + \tfrac{z}{2}) \gamma^+ W( y
+ \tfrac{z}{2}, y - \tfrac{z}{2}) q ( y - \tfrac{z}{2}), \nonumber \\ 
\end{eqnarray} 
where the summation over color is implicit and the
flavor indices are omitted for brevity.  As the quark and antiquark
coordinates in the individual operators $\mathcal{O}_{j}$ in
Eq.~(\ref{eq:defsd}) are not split in the transverse direction, the
light-front gauge $A^-_a=0$ along straight line paths eliminates the
Wilson gauge link operators $W$, setting them to identity (the
splitting in $\vect{b}$ occurs between the color singlet bilinears,
which is innocuous). Thus the complications in maintaining the gauge
invariance and the choice of the Wilson line, encumbering
TMDs~\cite{Collins:2011zzd} and the corresponding QCD
evolution~\cite{Vladimirov:2015fea,Echevarria:2015usa} (for a review
see, e.g., ~\cite{Scimemi:2019mlf} and references therein), do not occur
in the case of dPDFs.
                  
One may pass to the momentum representation via the Fourier transform
\begin{eqnarray}
D_{j_1 j_2}(x_1, x_2,\vect{q}) = \int d^2 \vect{b} \, e^{i \vect{b}\cdot\vect{q}}  D_{j_1 j_2}(x_1, x_2, \vect{b}), \label{eq:avy}
\end{eqnarray}
where the same symbol $D$ is used, with the argument distinguishing
the function from its transform.  The double distributions for the case
where $\vect{q}=\vect{0}$, denoted for brevity as 
\begin{eqnarray}
D_{j_1 j_2}(x_1,x_2) = D_{j_1 j_2}(x_1, x_2,\vect{q}=\vect{0})
\end{eqnarray}
acquire a special significance, as they satisfy the Gaunt-Stirling
(GS) sum rules~\cite{Gaunt:2009re}. These identities follow
straightforwardly~\cite{Gaunt:2012tfk} when the decomposition of the
parton operators in a basis of the light-front wave functions is made,
and are essentially statements on completeness of the Fock states as well
as flavor and longitudinal momentum conservation.  The sum rules read
\begin{eqnarray}
&&\sum_{i} \int_0^{1-x_2} \!\!\! dx_1 \,x_1 D_{ij}(x_1,x_2)=(1-x_2) D_{j}(x_2), \nonumber \\
&&\int_0^{1-x_2} \!\!\!\!\!\!\!\! dx_1 \, D_{i_{\rm val} j}(x_1,x_2) = 
(N_{i_{\rm val}}\!-\!\delta_{ij}\!+\!\delta_{\bar{i}j})D_j(x_2),
\end{eqnarray}
where $i_{\rm val}$ indicates the difference of the parton ($i$) and antiparton ($\bar{i}$) distributions, and 
\begin{eqnarray}
N_{i_{\rm val}}=\int_0^1 dx \, D_{i_{\rm val}}(x).
\end{eqnarray}

To satisfy the GS sum rules, we have argued in~\cite{Broniowski:2013xba,Broniowski:2016trx} that a practical approach is the top-down method, where 
one starts with  $n$-body parton distributions with the longitudinal momentum fractions constrained with \mbox{$\delta(1-x_1-...-x_n)$}, 
and subsequently generates distributions 
with a lower number of partons via marginal projections. This avoids complications of bottom-up attempts in constructing dPDFs from known
sPDFs, which cannot be unique and where one encounters problems~\cite{Gaunt:2009re,Golec-Biernat:2014bva}. Also, note a recent study~\cite{Diehl:2018kgr}
devoted to a verification of the GS sum rules in covariant perturbation theory and in light-cone perturbation theory.

The above discussion presumes a probabilistic interpretation of
dPDFs, similarly to sPDFs.  As has been known, however (see,
e.g.,~\cite{Kasemets:2014yna}), the need for renormalization of the
bare definitions (\ref{eq:defsd}) may in principle invalidate
positivity, as it inherently involves subtraction.  The issue is
subtle, as violation of positivity precludes a probabilistic
interpretation.  Moreover, it should also be recalled that the QCD
evolution equations (see Sec.~\ref{sec:evol} below) do not preserve
positivity; whereas the DGLAP evolution produces {\it positive}
distributions when evolving upward from $Q_0$ to $Q>Q_0$, for downward
evolution, $Q < Q_0$, the positivity property may actually be
violated~\cite{LlewellynSmith:1978me} (mostly at small $x$).  For an
explicit nucleon study and a practical distinction between the
(scheme dependent) positivity of parton distributions and the
necessary positivity of physically measurable cross sections, we
refer the reader to~\cite{RuizArriola:1998er}. The generation of negative
components with downward evolution is also a feature of the dPDF
evolution equations applied here.

Secondly, there is a question of retaining positivity at nonzero
$\vect{b}$, or a corresponding Fourier conjugate momentum $\vect{q}$.
The mentioned derivation of the GS sum rules for $\vect{q}=\vect{0}$
via the Fock-state decomposition involves a summation of moduli
squared of the $n$-parton wave functions,
$|\phi_n(\{\vect{k}_i\},\{x_i\})|^2$, which is positive and may remain
so upon certain renormalization procedures, e.g., the Fock space
truncation. With $\vect{q} \neq \vect{0}$, the corresponding terms
involve ${\rm Re} \left
[\phi_n(\{\vect{k}_i\},\{x_i\})\phi_n^\ast(\{\vect{k}_i+\vect{q}_i\},\{x_i\}) \right
]$, which mathematically need not be positive definite, as the wave
functions may possess nodes. Of course, in processes where hard-soft
factorization holds, thus where the dPDFs are useful, the pertinent
cross sections must obviously be positive, which constitutes the true
positivity constraints. The issue of positivity is further discussed in Sec.~\ref{sec:pos}.

We note that the factorization of the  longitudinal and transverse
degrees of freedom, 
\begin{eqnarray}
D_{j_1 j_2}(x_1, x_2, \vect{q}) = D_{j_1 j_2}(x_1, x_2) F_{j_1 j_2}(\vect q),
\end{eqnarray}
has been a standard working assumption in studies of DPS
over the recent years, which so far finds support from dynamical model
calculations~\cite{Chang:2012nw,Rinaldi:2013vpa}. We remark that in this case the positivity property can be considered 
separately for $D_{j_1 j_2}(x_1, x_2)$ and for the form factor $F_{j_1 j_2}(\vect q)$.

We end this section with remarks concerning the lattice QCD results of~\cite{Bali:2018nde,Zimmermann:2017ctb} in the context of our studies. There, the basic 
object is the correlation function, which bares some similarity to the dPDF definition (\ref{eq:defsd}), but also carries relevant differences. Firstly, typical of the lattice 
simulations, the currents are local, with $z_1^-$ and $z_2^-$ set to zero, which corresponds to the integration over $x_1$ and $x_2$. Second, the 
relative time difference of the two currents in~\cite{Bali:2018nde,Zimmermann:2017ctb} is set to zero, whereas in Eq.~(\ref{eq:defsd}) integration over 
$y^-$ is carried out. To focus more precisely on this issue, let us denote the relevant matrix element in terms of Lorentz invariants, 
\begin{eqnarray}
\langle p |\, {\mathcal{O}_{j_1}(y,0)\, \mathcal{O}_{j_2}(0,0)}\,| p \rangle = f _{j_1 j_2}(p\cdot y, y^2).
\end{eqnarray}
In the light cone kinematics of Eq.~(\ref{eq:defsd}) $p\cdot y= p^+
y^-$, $y^2=-\vect{b}^2$, and integration over $p\cdot y$ is carried
out. In the instant form of the lattice studies $p\cdot y= p^0 y^0$,
$y^2=-\vect{y}^2$, and $p\cdot y=0$. If, instead of integration over
$y^-$ in Eq.~(\ref{eq:defsd}) we set $y^-=0$, the Lorenz covariance
would immediately link the two approaches. Since it is not so, the
transverse form factors from dPDFs and the lattice evaluations are
different objects. For that reason the transverse form factors related
to dPDFs discussed in Sec.~\ref{sec:trans} cannot be directly compared
to the results presented in~\cite{Bali:2018nde,Zimmermann:2017ctb}. A
model interpretation of the interesting lattice data requires a
separate study which should also incorporate in addition the
corresponding finite pion mass, typically $\sim 300$~MeV, used in
those simulations.

\section{The model \label{sec:model}}

\begin{figure}
\begin{center}
\includegraphics[angle=0,width=0.25\textwidth]{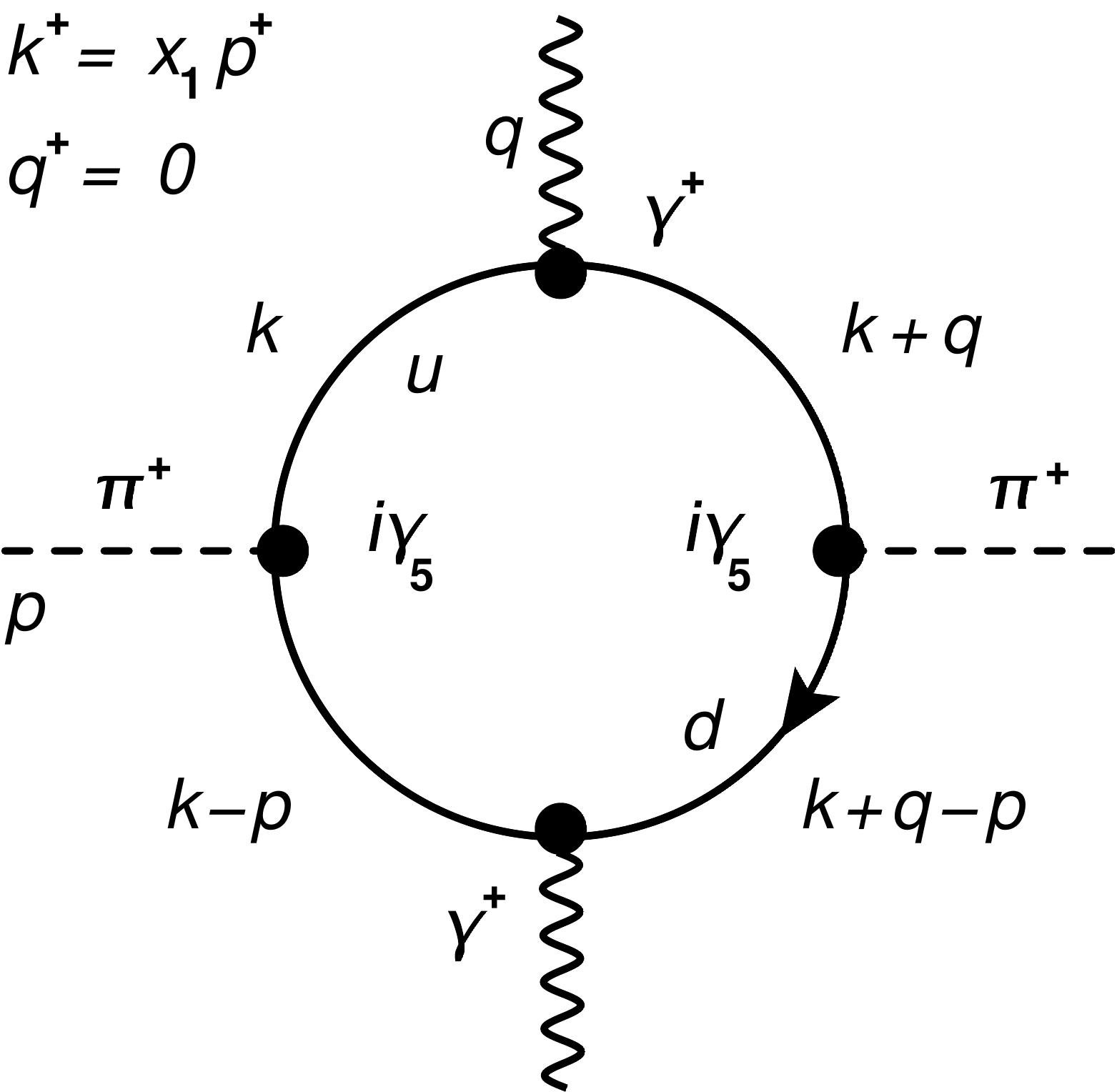}
\end{center}
\vspace{-4mm}
\caption{The diagram for the chiral quark model evaluation of the valence dPDF of $\pi^+$. The loop 
consists of the constituent quark propagators. The blobs 
indicate the probing operators $\gamma^+$, and the pseudoscalar quark-pion coupling $i \gamma_5$.
The integration over $\int dk_- \int d^2 k_\perp \int dq_-$ enforces the kinematic constraints of the definition~(\ref{eq:defsd}).
\label{fig:diag}} 
\end{figure}  

The large-$N_c$ evaluation in chiral quark models amounts to
evaluating one-quark-loop integrals.  For definiteness, we consider
the positively charged pion $\pi^+$, with other states related by the
isospin symmetry.  The relevant diagram for the valence dPDF 
in the momentum representation is shown in
Fig.~\ref{fig:diag}, where the loop momentum integration is
constrained with $k^+=x_1 p^+$, as we assign $x_1$ as the longitudinal
momentum fraction carried by the valence $u$ quark.  Note that with
just two constituents this constraint simultaneously fixes the
longitudinal momentum carried by the valence antiquark $\bar{d}$ to be
$x_2=1-x_1$. With the Feynman quark propagator
\begin{eqnarray}
&& S_k=\frac{i}{\slashed{k} -M+i\epsilon}, \label{eq:not} 
\end{eqnarray}
where $M$ is the constituent quark mass due to the spontaneous breaking of the chiral symmetry, 
the diagram of Fig.~\ref{fig:diag} reads
\begin{eqnarray}
&& \hspace{-2mm} D_{u \bar{d}}(x_1,x_2,\vect{q})=\delta(1-x_1-x_2) \frac{M^2}{f^2} \int \frac{d^4k}{(2\pi)^4} \int \frac{dq^-}{2\pi} \times \nonumber \\
&& \delta(k^+-x_1 p^+) {\rm Tr}\left [ \gamma^+ S_k \gamma^5 S_{k-p} \gamma^+ S_{k-p} \gamma_5 S_k \right],
\end{eqnarray}
where the trace is over color and Dirac indices.  We use the
pseudoscalar quark-pion coupling $M/f \times i\gamma_5$, where $f=86$~MeV is
the pion decay constant in the chiral limit.  An explicit and
straightforward evaluation yields~\cite{BW-ERA-LC2019,Courtoy:2019cxq}
\begin{eqnarray}
&& \hspace{-2mm} D_{u \bar{d}}(x_1,x_2,\vect{q})=\delta(1-x_1-x_2) \theta(x_1)\theta(1-x_1) F(\vect{q}), \nonumber \\ \label{eq:fact}
\end{eqnarray}
with the form factor
\begin{eqnarray}
\vspace{-10mm}F(\vect{q}) = \left . \frac{N_c M^2}{(2\pi)^3 f^2} \int d^2\vect{k} 
\frac{\vect{k}\cdot (\vect{k}+\vect{q})+M^2}{(\vect{k}^2+M^2)((\vect{k}+\vect{q})^2+M^2)} \right |_{\rm reg.}, \nonumber \\
\label{eq:ff-unreg}
\end{eqnarray}
which is ultraviolet log-divergent and needs to be properly regularized, as indicated. 

For $\vect{q}=\vect{0}$, upon regularization, the integral reduces to 
\begin{eqnarray}
F(0)= \left . \frac{N_c M^2}{(2\pi)^3 f^2} \int d^2\vect{k} 
\frac{1}{\vect{k}^2+M^2} \right |_{\rm reg.},
\label{eq:norm-cond}
\end{eqnarray}
which is equal to 1, as follows from a corresponding expression for the square of the pion decay constant $f^2$
(cf. for instance~\cite{Broniowski:2007si}). Then 
\begin{eqnarray}
&& \hspace{-2mm} D_{u \bar{d}}(x_1,x_2,\vect{q})=\delta(1-x_1-x_2) \theta(x_1)\theta(1-x_1), \label{eq:dPDF1}
\end{eqnarray}
the result advocated in \cite{BW-ERA-LC2019,Courtoy:2019cxq}.

As shown in \cite{Courtoy:2019cxq}, other large-$N_c$ diagrams, which formally appear in the
effective quark-meson model, do not contribute to $ D_{u \bar{d}}$.

Thus, at the quark model scale $\mu_0$, the double parton distribution
of the pion takes a product form of the individual longitudinal
momentum fractions carried by each valence quark, multiplied by
$\delta(1-x_1-x_2)$, coming from the momentum conservation and the fact
that at the quark model scale the only constituents in the pion are
the two valence partons.

We recall that for the valence sPDF the corresponding result
is~\cite{Davidson:1994uv}
\begin{eqnarray}
D_{\rm val}(x)=\theta(x)\theta(1-x). \label{eq:s}
\end{eqnarray}

Taking into account the fact that our model result factorize according
to Eq.~(\ref{eq:fact}), we discuss separately the longitudinal and
transverse structures in the following sections.

\section{Regularization vs positivity \label{sec:pos}}
\label{sec:reg-vs-pos}

In this section we address some technical aspects concerning the
interplay between the necessary finite cut-off regularization
in a chiral quark model and the positivity which proves the basis for
the insightful probabilistic interpretation. As already mentioned,
even in QCD this is a subtle issue~\cite{Kasemets:2014yna}. To
simplify the discussion, in this section we assume sharp
cut-offs either in coordinate or momentum space. 
Generally, these schemes encounter difficulties with gauge invariance. More sophisticated (smooth but gauge invariant) schemes
are worked out in Sec.~\ref{sec:trans}.

On a formal level, it is noteworthy that Eq.~(\ref{eq:ff-unreg}) has a
convolution-like structure that can be rewritten in the  form  
\begin{eqnarray}
\hspace{-5mm}F(\vect{q})&=& \frac{N_c M^2}{(2\pi)^3 f^2}
\int d^2 \vect{k}_1 d^2 \vect{k}_2 \delta(\vect{k_1}-\vect{k_2}+ \vect{q}) \nonumber \\
&\times &\left[ \Psi (\vect{k_1}) \vect{k_1} \cdot \Psi
(\vect{k_2}) \vect{k_2} + M^2 \Psi (\vect{k_1}) \Psi
(\vect{k_2}) \right] ,
\end{eqnarray}
where
\begin{eqnarray}
\Psi (\vect{k}) = \frac{1}{\sqrt{\vect{k}^2+M^2}}.
\end{eqnarray}
Formally, we may pass to the coordinate space by defining
\begin{eqnarray}
\Phi (\vect{b}) = \int \frac{d^2 \vect{k}}{(2\pi)^2} \Psi (\vect{k}) e^{i \vect{k} \cdot \vect{b}} =
\frac{e^{-bM}}{2\pi b},
\end{eqnarray}
with $\vect{b}$ the two-dimensional transverse coordinate. From here
we get
\begin{eqnarray}
\hspace{-5mm}F(\vect{q})=  \frac{N_c M^2}{2\pi f^2} \int d^2 \vect{b} \, e^{i \vect{b} \vect{q}}
\left[ \nabla \Phi (\vect{b})^2 + M^2 \Phi (\vect{b})^2 \right] ,
\end{eqnarray}
such that the form factor appears as the Fourier transform of a
manifestly positive function in $b$-space. In our case $\Phi
(\vect{b}) \sim e^{-M b}/b$, hence the integral diverges at small
distances. If we put a short distance transverse cut-off, it is not
guaranteed that the form factor in momentum space remains
positive.\footnote{Quite generally, the Fourier transformation of a
positive function is not necessarily positive.} In fact, if we impose
the normalization condition, $F(0)=1$, we get a short distance cut-off
$b_0 \sim 0.5~{\rm fm}$ and the form factor presents a multinodal
structure with alternating signs, the first zero curring at $
|\vect{q}| \sim 0.75~{\rm MeV}$. 

Alternatively, since the formula for $F(\vect{q})$ is divergent, a
regularization {\em must} be imposed. A simple method within
momentum space would be to use a purely transverse cut-off
$\Lambda_\perp$ and fix it according to the normalization condition
given by Eq.~(\ref{eq:norm-cond}).  This is the essence of the light
front regularization method applied in Ref.~\cite{Courtoy:2019cxq},
which, requires additional assumptions to be compatible with soft
physics and, in particular, to provide a non-vanishing vacuum quark
condensate (see e.g.~\cite{Dietmaier:1989hv,Heinzl:2000ht}). Such a
prescription, however, does not preserve
positivity~\cite{Courtoy:2019cxq}.

An interesting aspect is that, as noted in
\cite{Weigel:1999pc} (see Appendix A for an analysis in the case of the pion em form factor), usual convolution formulas,
as the one discussed above, are only {\it formally} correct but not
necessarily compatible with gauge invariance or chiral
symmetry.\footnote{As a matter of fact, rather than a simple
convolution one has {\it after} regularization a superposition of
convolutions with negative weights ensuring the finiteness of the
result (see \cite{Weigel:1999pc} and bellow).}

The above discussion shows once again that implementation of a finite
cut-off regularization in chiral quark models is a nontrivial issue,
particularly if a {\it finite} regularization is imposed separately on
the individual Feynman diagrams. The effective action method
suggested by Eguchi~\cite{Eguchi:1976iz} is a symmetry preserving
scheme which allows for a proper discussion of chiral symmetry
breaking even after the implementation of a regularization
method~\cite{Schuren:1991sc,RuizArriola:2002wr} and provides a {\it
common} regularization for all diagrams involving pions, photons, or
W and Z bosons.  Therefore, here we only consider regularization
methods which comply with soft pion physics and chiral symmetry, but
then they need not preserve positivity of the dPDF form factor away
from the soft limit (particularly above the finite cut-off). Thus,
chiral quark models are designed to describe soft physics, whereas
large $|\vect{q}|$, or small $b$, evade this requirement, even after
regularization, which we encounter below.

\section{Evolution and matching \label{sec:evol}}

Parton distribution functions describe non-perturbative properties of
a hadron, namely its quark and gluon content, as functions of their
longitudinal momentum fraction. These quantities depend on the
renormalization scale, which usually is taken to be $\mu = Q$, a
typical momentum appearing in the experimental process. Their running
with the scale $\mu$ follows from the invariance of the physical cross
sections. The evolution can only be implemented perturbatively, thus a
non-perturbative input is required at a reference scale $\mu_0$ to
provide initial conditions for the pertinent evolution equations.

Phenomenological analyses of the pion parameterize the 
valence, sea, and gluon sPDFs at relatively large scales, $\mu \sim 2~{\rm GeV}$~\cite{Sutton:1991ay}, 
or intermediate scales $\mu \sim 0.5~{\rm GeV}$~\cite{Gluck:1999xe}, with the result  that at 
$\mu \sim 2~{\rm GeV}$ the momentum fraction carried by the quarks in the pion is $0.42-0.44$.  
In fact, in a hadronic model such as NJL applied here, one only has valence
quarks and antiquarks, hence it makes sense to fix $\mu_0$ by imposing that
these constituents saturate the momentum sum rule, 
$\langle x \rangle_{\mu_0}=1$. This provides a
hadronic scale of $\mu_0 \sim 320~{\rm MeV}$ (see,
e.g.,~\cite{Broniowski:2016trx}), which we refer to as the {\em quark model scale}.
Once the reference scale $\mu_0$ is fixed, the matching condition between
the model and QCD observables is imposed by
\begin{equation}
\left . A(x,\mu_0) \right |_{\rm model} = \left . A(x,\mu_0) \right |_{\rm QCD}.
\end{equation}

The QCD evolution equations for multi-parton distributions have been
derived long ago~\cite{Kirschner:1979im,Shelest:1982dg}. A simple and
numerically efficient method is based on the Mellin moments, similarly
to the case of sPDFs (for results of a practical implementation see,
e.g.,~\cite{Broniowski:2013xba} for the nucleon and
\cite{Broniowski:2016trx} for the pion).

While for ease of notation we do not write explicitly the $\vect{b}$
dependence, the evolution equations quoted below are still
valid for that case~\cite{Diehl:2014vaa}. Clearly, for a factorized ansatz, the
transverse dependence also factorizes out from the evolution. As we
have discussed in the previous section, this is actually the case in
chiral quark models in the chiral limit.

One introduces the moments of the sPDFs and dPDFs,
\begin{eqnarray}
&& \hspace{-1mm} M^{n}_{j}=\int_0^1 dx \, x^n  D_j(x), \\
&& \hspace{-1mm} M^{n_1 n_2}_{j_1 j_2}=\int_0^1 \!\!dx_1 \int_0^1\!\! dx_2 \theta(1\!-\!x_1\!-\!x_2) x_1^{n_1} x_2^{n_2} D_{j_1 j_2}(x_1,x_2),  \nonumber
\end{eqnarray}
and the moments of the QCD splitting functions
\begin{eqnarray}
&& P_{i \to j}^n=\int_0^1 dx \, x^n P_{i \to j}(x), \\
&& P_{i \to j_1 j_2}^{n_1 n_2}=\int_0^1 dx \, x^{n_1} (1-x)^{n_2} P_{i \to j_1 j_2}(x), \nonumber \\
&& {\tilde{P}}_{i \to j_1 j_2}^{n_1 n_2}=\delta_{j_1 j_2}P_{i \to j_1}^{n_1+n_2} - 
 \delta_{i j_1}P_{j_1 \to j_2}^{n_2}  - \delta_{i j_2}P_{j_2 \to j_1}^{n_1}. \nonumber
\end{eqnarray}
Then the DGLAP evolution equations are \cite{Kirschner:1979im,Shelest:1982dg} (for simplicity, the possible two scales $\mu_1$ and $\mu_2$ 
for the two operators are set to be equal to a single scale $\mu$)
\begin{eqnarray}
&& \frac{d}{dt} M^{n_1 n_2}_{j_1 j_2} = \sum_i P_{i \to j_1}^{n_1} M^{n_1 n_2}_{i j_2} + \sum_i P_{i \to j_2}^{n_2} M^{n_1 n_2}_{j_1 i} \nonumber \\
&& ~~~+ \sum_i \left ( P_{i \to j_1 j_2}^{n_1 n_2} + {\tilde{P}}_{i \to j_1 j_2}^{n_1 n_2} \right) 
M^{n_1+n_2}_{i}, \label{eq:dDGLAP}
\end{eqnarray}
where 
\begin{eqnarray}
&& t= \frac{1}{2\pi \beta} \log \left [ 1+\alpha_s(\mu) \beta \log(\Lambda_{\rm QCD}/\mu) \right ], \\
&& \beta=\frac{11N_c-2N_f}{12\pi}.
\end{eqnarray}

Partons $i$, $j_1$, and $j_2$ may in general represent the valence quarks/antiquarks, the sea, or the gluons, whose 
distributions are coupled. Moreover, the last term in Eq.~(\ref{eq:dDGLAP}) couples dPDFs to sPDFs, serving as an 
inhomogeneous term. 
The evolution acquires a simpler form when one probes just the valence quarks,
which is the case considered in this  work, where we take $\pi^+$ with 
$j_1 = u$ and $j_2=\bar d$. In that case there are no partons $i$ contributing to the splitting $P_{i \to j_1 j_2}$, and the 
inhomogeneous term vanishes. In addition, the nonsiglet valence quarks do not mix with the singlet component (sea quarks and gluons),  
leaving for the considered $\pi^+$ state the equation 
\begin{eqnarray}
\frac{d}{dt} M^{n_1 n_2}_{u \bar d}(t) = \left ( P_{u \to u}^{n_1} + P_{\bar d \to \bar d}^{n_2} \right ) M^{n_1 n_2}_{u \bar d}(t).  \label{eq:pidD}
\end{eqnarray}
Figure~\ref{fig:cascade} represents a typical diagram for the applied evolution in 
terms of the developing parton cascades along the $t$-channel. 

\begin{figure}
\begin{center}
\includegraphics[angle=0,width=0.4 \textwidth]{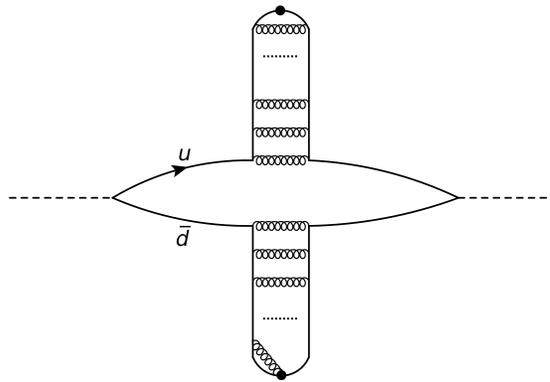} 
\end{center}
\vspace{-2mm}
\caption{A typical QCD diagram corresponding to the evolution (\ref{eq:pidD}) of the valence dPDF of $\pi^+$. The dots 
correspond to insertions of $\gamma^+$.
\label{fig:cascade}} 
\end{figure}

\section{Numerical results of evolution \label{sec:num}}

\begin{figure*}
\begin{center}
\includegraphics[width=0.39\textwidth]{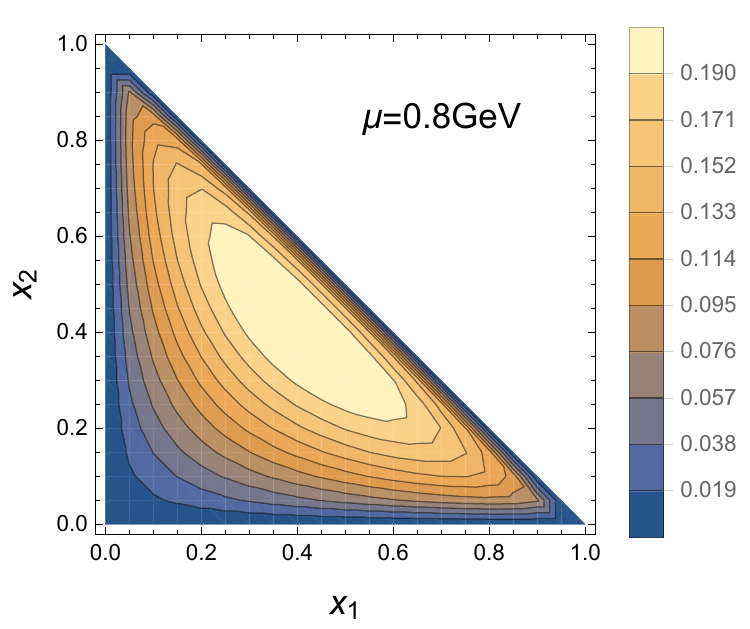}
\includegraphics[width=0.39\textwidth]{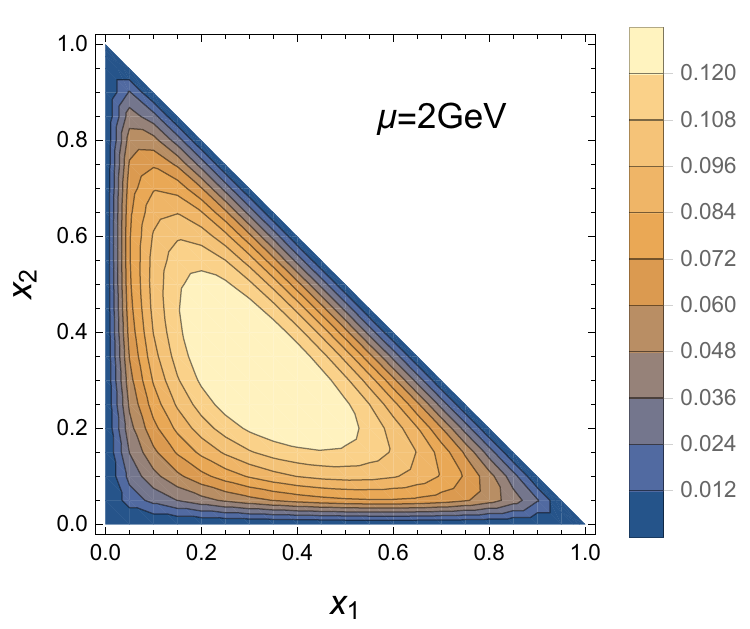} \\
\includegraphics[width=0.39\textwidth]{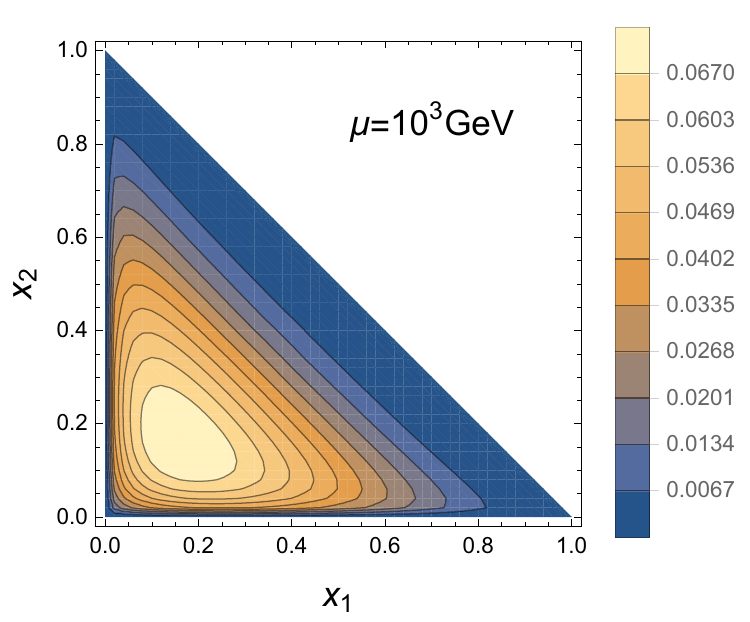}
\includegraphics[width=0.39\textwidth]{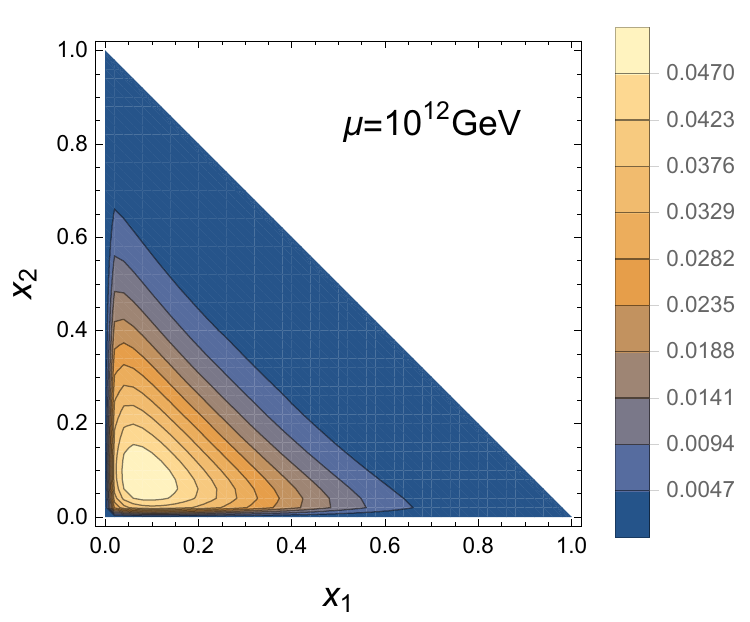}  
\caption{The valence dPDF of $\pi^+$ evolved with the DGLAP equations 
to subsequent scales of $\mu=0.5$, $2$, $10$, $1000$, and $10^9$~GeV. The dimensionless 
quantity plotted is $x_1 x_2 D(x_1,x_2;\mu)$. The initial condition at $\mu=\mu_0$ 
is the singular distribution of Eq.~(\ref{eq:dPDF1}). 
\label{fig:evol}} 
\end{center}
\end{figure*}  

The DGLAP evolution described above in an obvious manner preserves the longitudinal-transverse factorization, 
as all moments are multiplied conventionally with $F(\vect{q})$. For that reason 
we may discuss the evolution of $D_{\rm u \bar{d}}(x_1,x_2)$.

It is known that the DGLAP evolution preserves the GS sum
rules~\cite{Gaunt:2009re}. In our case, the sum rules are trivially satisfied at
the quark model scale $\mu_0$, as with Eqs.~(\ref{eq:fact}) and (\ref{eq:s})
one immediately gets
\begin{eqnarray}
&& \int_0^{1-x_2} dx_1 D_{\rm u \bar{d}}(x_1,x_2) = D_{\bar{d}}(x_2), \nonumber \\ && \int_0^{1-x_1} dx_2 D_{\rm u \bar{d}}(x_1,x_2) = D_{{u}}(x_1), \nonumber \\
&& \int_0^{1-x_2} dx_1 x_1 D_{\rm u \bar{d}}(x_1,x_2) = (1-x_2) D_{\bar{d}}(x_2), \nonumber \\  && \int_0^{1-x_1} dx_2 x_2 D_{\rm u \bar{d}}(x_1,x_2) = (1-x_1)D_{{u}}(x_1).
\end{eqnarray}
Thus the GS sum rules are satisfied at any scale. 

Our numerical results for the evolution of the valence dPDF of the pion 
(multiplied with $x_1 x_2$) are presented in Fig.~\ref{fig:evol}, where we show it for increasing
evolution scale $\mu$. We note the gradual drift towards the asymptotic  fixed point $\delta(x_1) \delta(x_2)$. 

\begin{figure*}
\begin{center}
\includegraphics[width=0.39\textwidth]{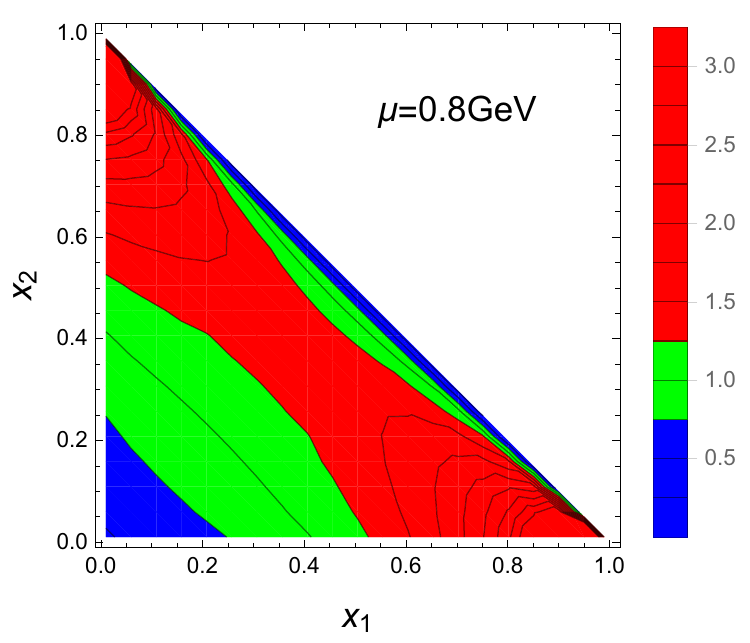}
\includegraphics[width=0.39\textwidth]{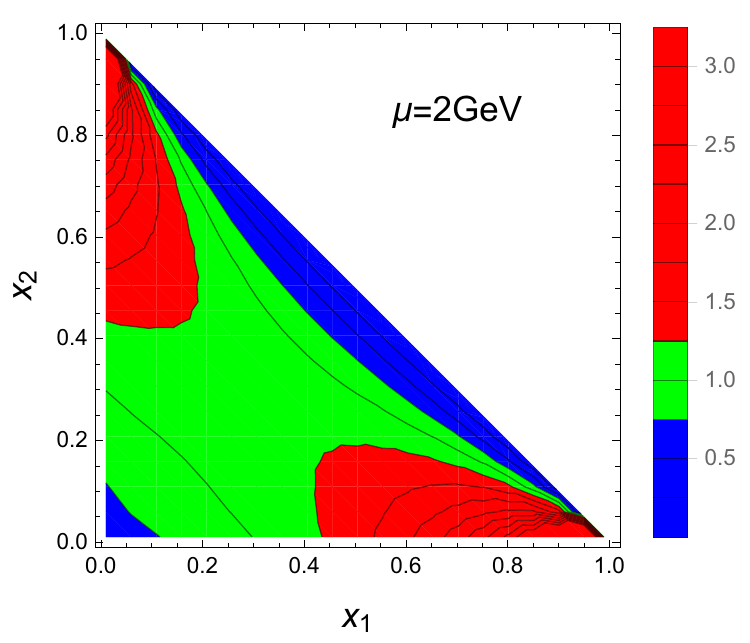} \\
\includegraphics[width=0.39\textwidth]{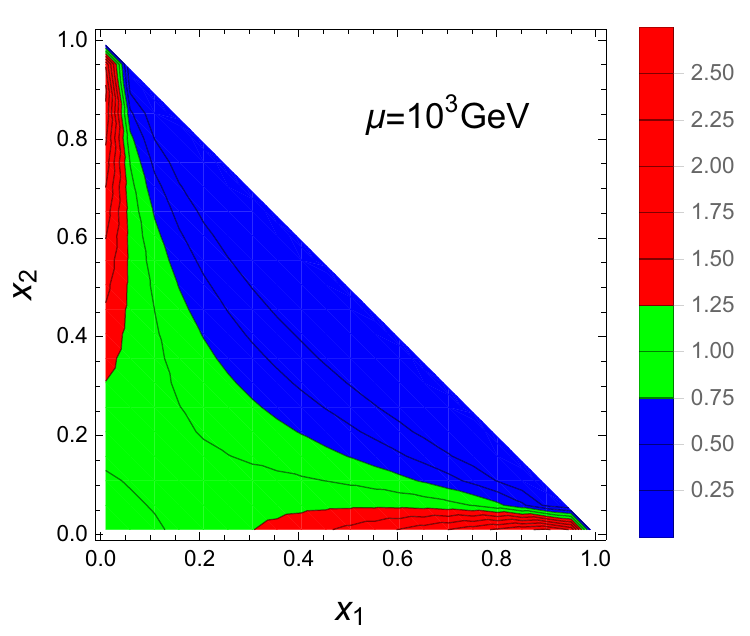}
\includegraphics[width=0.39\textwidth]{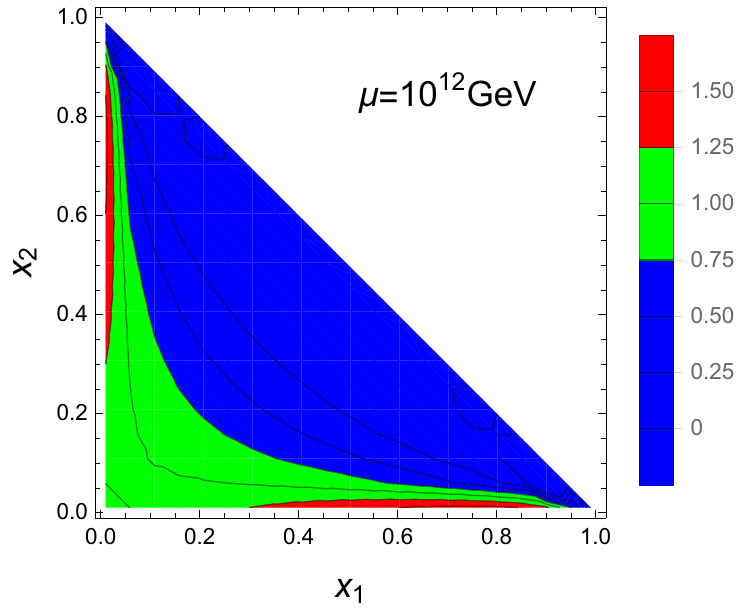}  
\caption{Correlation $D_{\rm u \bar{d}}(x_1,x_2)/D_u(x_1)D_{\bar{d}}(x_2)$ at various evolution scales $\mu$. 
\label{fig:corr}} 
\end{center}
\end{figure*}  

The longitudinal correlations can be conveniently quantified with the
ratio $D_{\rm u \bar{d}}(x_1,x_2)/D_u(x_1)D_{\bar{d}}(x_2)$, which
would be equal to 1 if no correlations were present. Our results for
this measure are shown in Fig.~\ref{fig:corr}. We note that for large
scales $\mu$ and simultaneously low $x_1$ and $x_2$ (which is where
dPDF is large) the correlation ratio is within a 20\% band around
unity, hence there the effect is not very substantial, justifying
the product ansatz for dPDF.  The largest effect occurs for asymmetric
kinematics, with $x_1$ large and $x_2$ small, or vice versa, where
correlations are strong and positive.  The explanation of this
behavior is related to the overall conservation of the longitudinal
momentum. The qualitative argument here is that at large $\mu$, with
many partons present, the constraint of the longitudinal momentum is
effective only when one of the considered partons takes a large
momentum fraction, thus leaving significantly less of available phase
space for the other parton.  We remark that a qualitatively similar
effect has been found for gluodynamics
in~\cite{Golec-Biernat:2015aza,Golec-Biernat:2016vbt,Elias:2017flu}.

\section{Ratios of moments \label{sec:ratios}}

The LO DGLAP evolution of the valence moments, due to the absence of
the inhomogeneous term, leads to a simple fact that the ratios
$\langle x_1^n x_2^m \rangle /\langle x_1^n \rangle \langle
x_2^m \rangle$ for the valence distributions do not depend on the
evolution scale, as the evolution ratio factors with the anomalous
dimensions cancel out.  
More explicitly, the evolution for the valence moments reads
\begin{eqnarray}
&& M_n(\mu)=\left (\frac{\alpha(\mu)}{\alpha(\mu_0)}\right )^{\gamma_n/2\beta_0}M_n(\mu_0), \nonumber \\
&& M_{nm}(\mu)=\left (\frac{\alpha(\mu)}{\alpha(\mu_0)}\right )^{\gamma_{n}/2\beta_0 + \gamma_{m}/2\beta_0 }M_{nm}(\mu_0), \label{eq:anom}
\end{eqnarray}
with $\gamma_i$ denoting the anomalous dimensions, thus the cancellation is obvious. 
Despite its simplicity, this feature is rather
remarkable, as it allows for an insight into lower scales, where
non-pertubative dynamics sets in, from the information at higher scales.

In particular, in the NJL model we immediately find the following scale independent ratios:
\begin{eqnarray}
\frac{\langle  x_1^n x_2^m \rangle}{\langle  x_1^n \rangle \langle x_2^m \rangle}= \frac{(1+n)!(1+m)!}{(1+n+m)!}. \label{eq:momnjl}
\end{eqnarray}
The first few ratios are shown in Table~\ref{tab:rat}. The largest ratio is for $m=n=1$.
Naturally, the moments are the quantities to be probed with the upcoming lattice studies and it is interesting to see if the pattern of 
Eq.~(\ref{eq:momnjl}) holds. The pattern of the moments is specific to a given (non-perturbative) model, allowing for its scrutiny
provided the lattice data would be accurate enough.

\begin{table}
\label{tab:rat}
\caption{The NJL model ratios of the valence moments  $\langle  x_1^n x_2^m \rangle /\langle  x_1^n \rangle \langle x_2^m \rangle$ from Eq.~(\ref{eq:momnjl}).
Rows and columns correspond to $n$ and $m$.}
\begin{tabular}{c|ccccc}
   & 1 & 2 & 3 & 4 & 5 \\ \hline
 1 & $\tfrac{2}{3}$ & $\tfrac{1}{2}$ & $\tfrac{2}{5}$ & $\tfrac{1}{3}$ & $\tfrac{2}{7}$ \\
 2 & $\tfrac{1}{2}$ & $\tfrac{3}{10}$ & $\tfrac{1}{5}$ & $\tfrac{1}{7}$ & $\tfrac{3}{28}$ \\
 3 & $\tfrac{2}{5}$ & $\tfrac{1}{5}$ & $\tfrac{4}{35}$ & $\tfrac{1}{14}$ & $\tfrac{1}{21}$ \\
 4 & $\tfrac{1}{3}$ & $\tfrac{1}{7}$ & $\tfrac{1}{14}$ & $\tfrac{5}{126}$ & $\tfrac{1}{42}$ \\
 5 & $\tfrac{2}{7}$ & $\tfrac{3}{28}$ & $\tfrac{1}{21}$ & $\tfrac{1}{42}$ & $\tfrac{1}{77}$ \\
\end{tabular}
\end{table}

\begin{figure}
\includegraphics[width=0.4\textwidth]{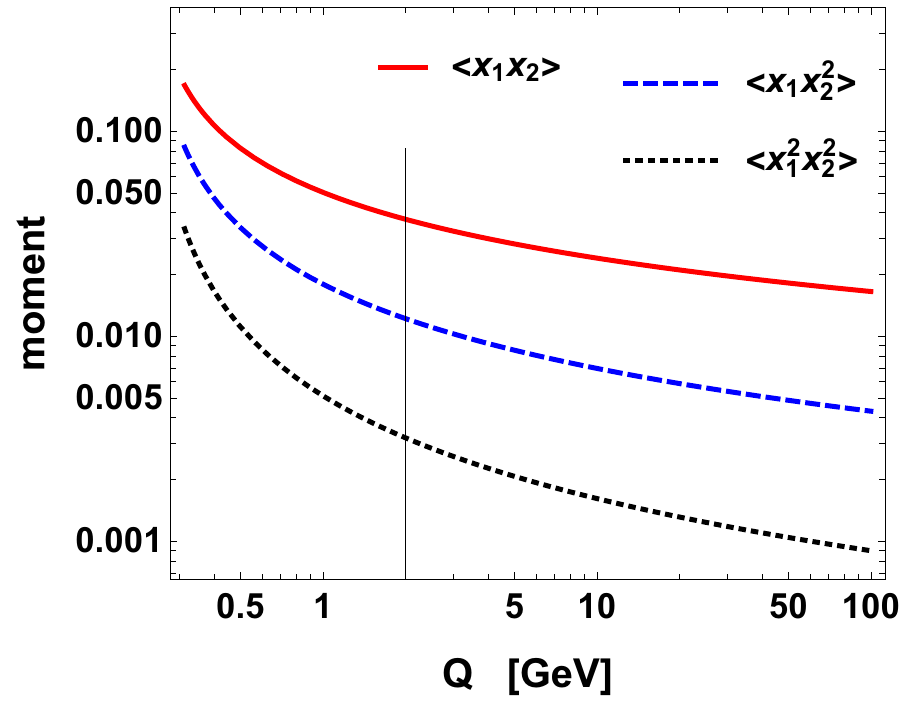} 
\caption{Lowest moments of the valence dPDF of the pion as a function of the renormalization scale according to LO DGLAP evolution. 
\label{fig:mom-evol}} 
\end{figure}

The LO DGLAP evolution of the lowest moments of the valence dPDF of
the pion is displayed in Fig.~\ref{fig:mom-evol}. We note a fall-off
from the quark model scale controlled by the corresponding anomalous
dimensions in Eq.~(\ref{eq:anom}).

\section{Transverse structure \label{sec:trans}}

The NJL model needs to be regularized to get rid of the ultraviolet
divergences, leaving only the soft-momentum degrees of freedom in the
dynamics. 
Unlike the longitudinal structure, the transverse structure depends
specifically on the regularization. This is a subtle issue, since as already discussed, the
chiral and gauge symmetries must be preserved by the regularization procedure, as has been
discussed at length in~\cite{Schuren:1991sc} (for a review see, e.g.,~\cite{RuizArriola:2002wr}). We give here the final recipes as
applied to the unregularized result of Eq.~(\ref{eq:ff-unreg}), both
for the PV and SQM implementations. We present first
the SQM results, as the final formulas are particularly simple in this case.

\subsection{Spectral Quark Model}

In the SQM, one replaces the constituent quark mass $M$ in
Eq.~(\ref{eq:ff-unreg}) with a {\it spectral mass},
$w$, and integrates over $w$ with a spectral weight,
\begin{eqnarray}
A(M)|_{\rm SQM} = \int_C d w \rho (w) A(w) 
\end{eqnarray}
over a suitably chosen complex
contour $C$~\cite{RuizArriola:2003bs}. One may adjust $\rho (w)$ and $C$ in such a way that the
construction implements an exact vector-meson dominance principle of the pion
electromagnetic form factor, namely 
$F_{\rm em} (Q^2) = 1/(1+Q^2/m_\rho^2)$, with $m_\rho^2 = 24
\pi^2 f^2 /N_c $, yielding $m_\rho= 764~{\rm MeV}$ at $f=86$~MeV in the chiral limit.
In this scheme, the normalization condition is $\int_C dw \rho(w)=1$. 

After computing the spectral integral as well as the $\vect{k}$
integral, the form factor is given by a very simple formula
\begin{eqnarray}
F(\vect{q}) = \frac{m_\rho^4-\vect{q}^2 m_\rho^2}{\left(m_\rho^2+\vect{q}^2\right){}^2}, \label{eq:FF}
\end{eqnarray}
which is a combination of the dipole (with positive sign) and monopole
(with negative sign) form factors. The form factor $F(\vect{q})$ is
correctly normalized at $\vect{q}=0$, namely $F(0)=1$, and vanishes as ${\cal
O} (|\vect{q}|^{-2})$ for large $|\vect{q}|$.

In Fig~\ref{fig:FF}(a) we show the form factor $F(\vect{q})$ 
for SQM (solid line).  As we can see, in accordance with Eq.~(\ref{eq:FF}),
this function becomes negative for $|\vect{q}| > m_\rho$. Thus the
result does nor obey positivity (cf. the discussion at the end of Sec.~\ref{sec:defs}). 
Passing to the configuration space via the Fourier-Bessel transform we get
\begin{eqnarray}
f(b) &=& \int \frac{d^2 \vect{q}}{(2\pi)^2} e^{i \vect b \cdot 
\vect{q}} F(\vect{q}) \nonumber \\ &=& \frac{m_\rho^2}{2\pi}\left[
b m_\rho K_1\left(b m_\rho\right)- K_0\left(b m_\rho \right) \right],
\end{eqnarray}
with $K_n(x)$ denoting the modified Bessel functions of order $n$. 

The form factor combination $b f(b)$ for SQM is 
is depicted in Fig~\ref{fig:FF}(b) with a solid line. By definition, this
function is normalized to unity, $\int d^2 b f(b) =1$, and the
corresponding mean squared radius (msr) is given by
\begin{eqnarray}
\langle b^2 \rangle_{\rm dPDF} = \int d^2 b\, b^2 f(b) = \frac{12}{m_\rho^2}=(0.88~{\rm fm})^2 .
\end{eqnarray}
This radius is a 2D quantity in relative parton-parton transverse coordinates,
and naively one expects a geometric relation to the 3D radius, which would
yield $\langle r^2 \rangle = \tfrac{3}{2} \langle b^2 \rangle  = (1.07~{\rm fm})^2$.  It is interesting to compare this measure with another msr
regarding the hadron size, for instance the electromagnetic (em) msr, which in SQM is given by
$\langle r^2 \rangle_{\rm em} = 6 /m_\rho^2 = \tfrac{1}{2} \langle b^2 \rangle_{\rm dPF}= (0.62~{\rm fm}^2) $. Thus, in this model the transverse size
(a 2D object) is about $\sqrt{2}$ larger than the em msr (a 3D
object).  At large distances $f(b)$ behaves as
\begin{eqnarray}
f(b) = \frac{\sqrt{b}\, m_\rho^{5/2} e^{-b m_\rho}}{2 \sqrt{2 \pi }} \left[ 1 + {\cal O} \left(\frac{1}{m_\rho b} \right) \right],
\end{eqnarray}
which falls off exponentially with the mass $m_\rho$, which in a
constituent picture is typically $m_\rho \simeq 2 M$ corresponding to
a double parton property, opposite to the $m_\rho/2 \simeq M$ scales
in single parton properties~\cite{RuizArriola:2003bs}.

At short distances $f(b)$ becomes negative and divergent, since
\begin{eqnarray}
f(b)= \frac{m_\rho^2}{2\pi} \left[ \log (\tfrac{1}{2} b m_\rho )+ \gamma +1  \right]+O\left(b^2\right),
\end{eqnarray}
where $\gamma=0.5772 \dots$ is the Euler-Mascheroni constant. Using
this asymptotics we get a zero of the function $f(b)$ for the value
$b_0 \sim 2 e^{-1-\gamma}/m_\rho \sim 0.1~{\rm fm}$. In Fig.~\ref{fig:FF}(b)
we show the $b-$dependence and, as we can see, the function becomes
negative at $b \sim 0.15~{\rm fm }$, which by the way is comparable to the
current day spacing of fine QCD lattices, $a \sim 0.1~{\rm fm}$.

\begin{figure}
\begin{center}
\includegraphics[angle=0,width=0.4\textwidth]{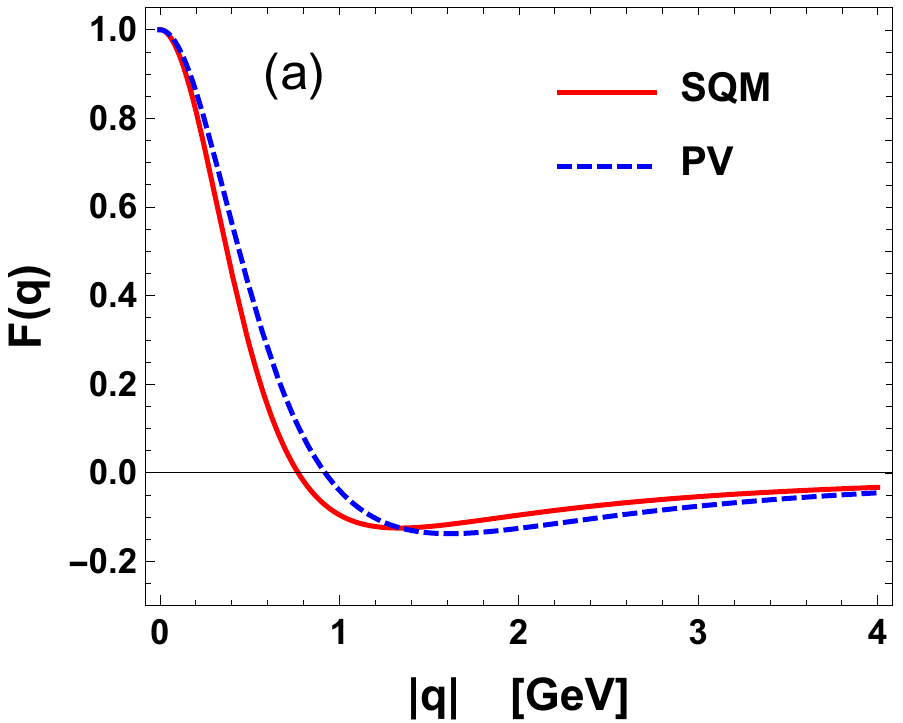} \\
\vspace{4mm}
\includegraphics[angle=0,width=0.4\textwidth]{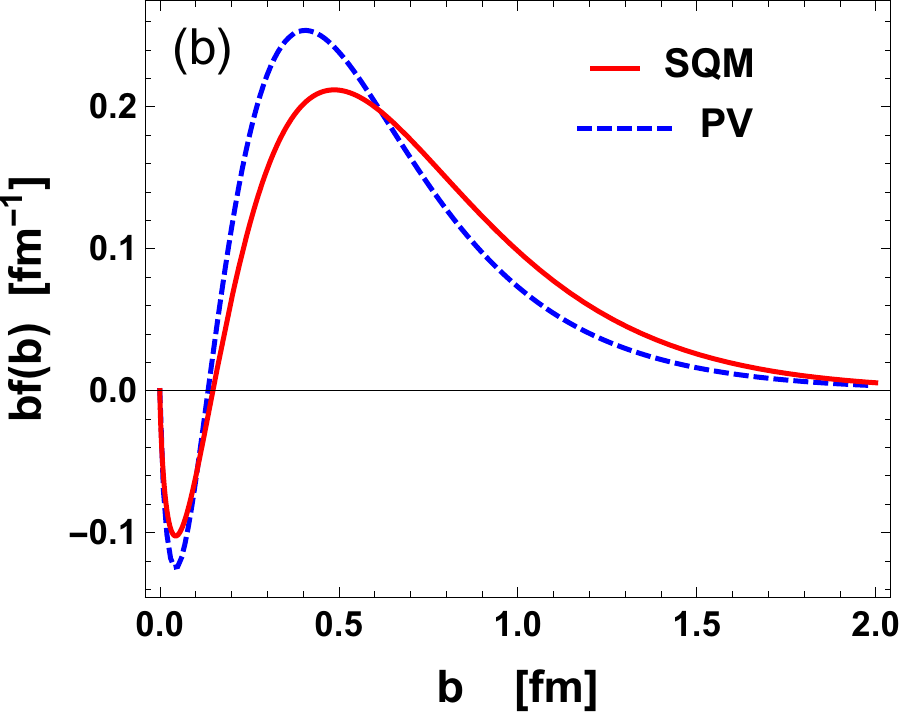} 
\end{center}
\vspace{-5mm}
\caption{Valence dPDF form factor of the pion in the Spectral Quark Model (solid lines) and in NJL with PV regularization (dashed lines), plotted
as functions of the transverse momentum (a) and the transverse coordinate (b), in the combination $b f(b)$. 
\label{fig:FF}} 
\end{figure}  

The effective cross section for DPS is a phenomenologically important quantity~\cite{DelFabbro:2000ds}, defined as
\begin{eqnarray}
\sigma_{\rm eff} = \frac{1}{\int \frac{d^2\vect{q}}{(2 \pi)^2} 
F (\vect{q}) F (-\vect{q}) }.
\end{eqnarray}
For SQM we get
\begin{eqnarray}
\sigma_{\rm eff} = \pi \frac{12}{m_\rho^2} = \pi \langle b^2 
\rangle_{\rm dPDF} = 23~{\rm mb},
\end{eqnarray}
which coincides {\it exactly} with the geometric cross section (see
the discussion in~\cite{Rinaldi:2013vpa}).  This number is
somewhat larger from similar estimates for the nucleon case, where one
obtains about 15~mb (see for instance~\cite{Blok:2011bu,Traini:2016jru,Blok:2013bpa})

Actually, if we restricted the integration limit to the value where the
effective form factor $f(b)$ is positive, we would get about twice
the value for the effective cross section, 47~mb.

Within this context there exists a nested bound for $\sigma_{\rm eff}$, found recently by Rinaldi and Ceccopieri~\cite{Rinaldi:2018slz}, namely
$ \pi \langle b^2 \rangle \le \sigma_{\rm eff} \le 3 \pi \langle b^2 \rangle $.  The
derivation of these interesting inequalities rests on the positivity of $f(b)$ 
(such that $\langle b^n \rangle \ge \langle b \rangle^n$). The lower bound is saturated by a monopole (which is
positive) as well as our solution (which becomes negative for
$|\vect{q}| \ge m_\rho $ or for $b \lesssim 0.5/m_\rho $). To compare, a dipole yields
$\sigma_{\rm eff} = \tfrac{3}{2}\pi \langle b^2 \rangle $, and a Gaussian profile
gives $\sigma_{\rm eff} = 2\pi \langle b^2 \rangle $.

\subsection{NJL with Pauli-Villars regularization}

The most straightforward regularization scheme satisfying the
necessary formal requirements is PV regularization with two
subtractions in the coincidence limit~\cite{Schuren:1991sc}, where the
one-quark-loop quantity $A(M^2)$ is replaced with
\begin{eqnarray}
A(M^2)|_{\rm reg, PV} &\equiv& A(M^2) - A(M^2\!+\!\Lambda^2) \nonumber \\
&+&  \Lambda^2 \frac{d}{d\Lambda^2}A(M^2\!+\!\Lambda^2). 
\end{eqnarray}
A detailed analysis shows that the rule is {\it only} applied to the
mass dependence under the integral in Eq.~(\ref{eq:ff-unreg}), but not
to the $M^2$ factor in front, originating from the coupling constant. The normalization yields
the well-known condition
\begin{eqnarray}
1 &=& \frac{N_c M^2}{(2\pi)^3 f^2} \int d^2\vect{k} 
\frac{1}{\vect{k}^2+M^2} \Big|_{\rm reg} \nonumber \\ 
&=& \frac{N_c M^2}{4 \pi^2 f^2} \left[ \log \frac{M^2+\Lambda^2}{M^2} + \frac{\Lambda^2}{\Lambda^2+M^2} \right] ,
\end{eqnarray}
linking the constituent quark mass and the PV cutoff $\Lambda$ at a
given value of $f$.  Following earlier works, we take $M=300~{\rm
MeV}$ and $f=86~{\rm MeV}$ in the chiral limit, which fixes the PV
cut-off to be $\Lambda=731~{\rm MeV}$. The expression for the valence
dPDF form factor is
\begin{eqnarray}
F(\vect{q}) &=& \frac{M^2 N_c}{4 \pi^2 f^2} \Big[ -\log (M^2)+
\frac{|\vect{q}|}{\sqrt{4 M^2+\vect{q}^2}} \nonumber \\
&\times& \log \left(  \frac{\sqrt{4 M^2+\vect{q}^2}-|\vect{q}|}{\sqrt{4 M^2+\vect{q}^2}+|\vect{q}|} \right) \Big] \Big|_{\rm reg.} .
\end{eqnarray}
Its asymptotic behavior is similar to SQM. Moreover,  
the msr is given by 
\begin{eqnarray}
\langle b^2 \rangle =\frac{N_c \Lambda ^4}{2 \pi ^2 f^2 \left(M^2+\Lambda ^2\right)^2} = (0.77~{\rm fm)^2}
\end{eqnarray}
and the effective cross section is numerically close to the geometric
one, $\sigma_{\rm eff} \sim \pi \langle b^2 \rangle $. The
corresponding form factors in the momentum and configuration spaces can
also be seen in Fig.~\ref{fig:FF} (dashed lines). As we can appreciate, they are very
similar to the SQM case, including negative values at high
momenta $\sim 1~{\rm GeV}$ or, equivalently, at short transverse
distances $b \lesssim 0.1~{\rm fm}$.

\section{Conclusions \label{sec:concl}}

We have presented a detailed study of the valence double parton distributions in the pion
in chiral quark models, with the following basic results:

\begin{itemize}

 \item The model leads in the chiral limit to a factorization of the
 longitudinal and transverse structure of the valence pion dPDF, in
 accordance to the findings
 of~\cite{BW-ERA-LC2019,Courtoy:2019cxq}. At the quark model scale,
 the longitudinal distribution is of the simple form
 $\delta(1-x_1-x_2)$, reflecting the momentum conservation with just
 two constituents.
 
 \item The LO DGLAP evolution is carried out in the Mellin space, leading
 to radiative generation of partons at higher scales, and resulting with 
 distributions at higher scales, accessible in experiments or lattice simulations.
 
 \item The Gaunt-Stirling sum rules are explicitly satisfied in our approach.
 
 \item The longitudinal correlations, quantified as the departure of dPDF
 from the product of two sPDFs, are studied in detail.  An assessment
 of the validity of the frequently used product ansatz is made, with
 the result that at high evolution scales and small momentum fractions
 of both constituents it works to a good approximation.
 
 \item Simple expressions are obtained for the transverse form factor
 of the valence dPDF in the applied regularizations of the chiral
 quark models. The issue of regularization and positivity is
 discussed. 
 
 \item Specific ratios of moments of valence dPDF to product of
 moments of valence sPDFs follow from the applied model. Such
 quantities, invariant of the evolution scale, may be used with future
 lattice data to scrutinize non-perturbative models of the pion
 structure.
 
\end{itemize}

\acknowledgments

This research was supported by the Polish National Science Centre (NCN)
Grant 2018/31/B/ST2/01022, the Spanish Ministerio de Economia y
Competitividad and European FEDER funds (grant FIS2017-85053-C2-1-P)
and Junta de Andaluc\'{\i}a grant FQM-225.

\bibliography{refs,dPDF}

\end{document}